\documentstyle[prb,aps,multicol,epsf]{revtex}

\begin{document}
\draft

\title{Evidence for an angular dependent contribution from columnar
defects to the equilibrium magnetization of YBa$_2$Cu$_3$O$_{7-\delta}$
}
\author{B. Hayani}
\address{Facult\'{e} des Sciences Dhar-Mehraz, B.P. 1796,
F\'{e}s-Atlas, Morocco}
\author{S. T. Johnson, L. Fruchter}
\address{Laboratoire de Physique des
Solides, B\^{a}timent 510, Universit\'{e} Paris-Sud, CNRS,
91405 Orsay, France.
}
\author{C. J. van der Beek}
\address{Laboratoire de Solides Irradi\'{e}s, CNRS UMR--7642,
\'{E}cole Polyt\'{e}chnique, 91128 Palaiseau, France.
}

\date{\today}
\maketitle

\begin{abstract}

We have measured an angle--dependent contribution to the
equilibrium magnetization of a YBa$_2$Cu$_3$O$_{7-\delta}$
single crystal with columnar defects created by irradiation with 5.8 GeV Pb
ions.
This contribution manifests itself
as a jump in the equilibrium torque signal, when the magnetic field
direction crosses that of the defects. The magnitude of the jump, which is
observed in a narrow temperature interval of less than 2 K wide, for
fields up to about twice the dose equivalent field $B_{\phi}$, is used to
estimate the energy gained by vortex pinning on the defects.  The
vanishing  of the effective
pinning energy at a temperature below $T_{c}$ is attributed to its
renormalization by thermal
fluctuations.

\end{abstract}


\begin{multicols}{2}
\narrowtext

\section{Introduction}

Since the discovery of increased pinning in heavy-ion irradiated samples,
the interaction between flux-lines and columnar defects in
high-T$_c$ superconductors has been the subject of intense
experimental and theoretical investigations.\cite{1,2}
An angle--dependent critical current enhancement has been put into
evidence both in the moderately anisotropic material
YBa$_2$Cu$_3$O$_{7-\delta}$ \cite{1} as in highly anisotropic
materials such as Bi$_2$Sr$_2$CaCu$_2$O$_{8+\delta}$.\cite{3,4}
The influence of correlated disorder on the equilibrium properties of the
flux-line lattice has been the subject of fewer experimental investigations.
Torque and magnetization experiments on  Bi$_2$Sr$_2$CaCu$_2$O$_{8+\delta}$
have revealed a pinning energy contribution to the equilibrium
magnetization arising from the presence of the defects, but,
in contrast to the irreversible magnetic moment, the equilibrium
magnetization did not show any other angle--dependent contribution
than the one arising from the layering of
the material.\cite{5} Here, we investigate the less anisotropic
compound YBa$_2$Cu$_3$O$_{7-\delta}$ and show that there exists a narrow
domain in the (H-T)
diagram where a reversible angular dependent contribution to the
torque arises due to the interaction of flux-lines with the linear defects.
This
constitutes a direct demonstration that vortex lines in the liquid phase
distort
in order to accomodate to the linear irradiation defects.

\section{Experimental details}

The experiments were performed on a single crystal of dimensions
$130 \times 37 \times 18$ $\mu$m$^{3}$; the shortest dimension was along
the $c$--axis. The transition temperature after irradiation was
$T_{c}=90.3$ K.
A single domain of parallel twin planes was observed, the planes
running at 45$^{\circ}$ with respect to the crystal's longest edge.
The sample was irradiated with 5.8 GeV Pb ions to a dose
$n = 10^{11}$ cm$^{-2}$, equivalent to a matching
(dose equivalent) field $B_{\Phi}=\Phi_{0} n = 20$ kG. The ion beam
was directed perpendicular to the longer crystal edge, at an
angle of 30 degrees with respect to the $c$-axis. The irradiation
created continuous linear amorphous defects of radius $c_{0} = 35$
\AA, oriented along the direction of the ion beam,\cite{12} with density $n$.
After the irradiation, the sample was characterized
using the magneto-optic flux visualization technique at 65, 77 and 82 K;
no evidence of any remaining influence of the twin planes on the flux
penetration
could be observed.

\begin{figure}
\centerline{\epsfxsize=9.0cm\epsfbox{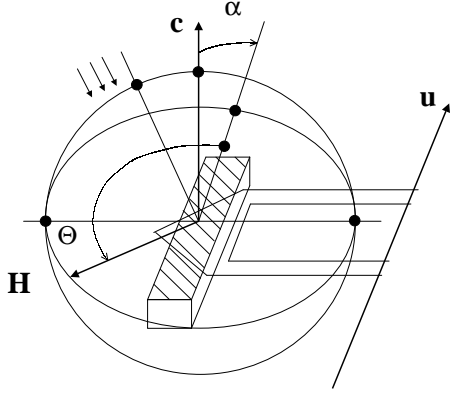}}
\caption{Scheme depicting the torque microlever and the orientation of
the applied field, the crystalline $c$--axis, and the direction of the
columnar defects.
}
\label{lever}
\end{figure}

The torque was measured using a piezoresistive microlever from Park
Scientific Instruments, as described in Ref.~\onlinecite{6}. The microlever
formed part of a low temperature Wheatstone resistive bridge, in
which a second lever with no sample was inserted in order to
compensate for the background signal originating from the
magnetoresistance of the levers. The measuring lever was fed with a current
of 300 $\mu$A and thermalized to better than 0.01 K using He${^4}$
exchange gas. The torque setup was calibrated from the Meissner
slope of the reversible magnetization as a function of field at a fixed
angle, as described elsewhere\cite{6}. In torque experiments with a
single rotation axis, the plane in which the applied field $H$ is
rotated is always at a misorientation angle $\alpha$ with respect to the plane
enclosing the $c$-axis and the irradiation direction (Fig.~\ref{lever}).
This angle is not known {\it a priori}: it results from  the uncertainty
in both the irradiation direction and the sample positioning. We estimate
that it
is less than a few degrees. The result of the misorientation is that the
applied field
is never strictly aligned with the irradiation direction; $\alpha$ is
therefore the minimum angle between $H$ and the ion tracks when the
field direction is varied.

\begin{figure}
\centerline{\epsfxsize=9.0cm\epsfbox{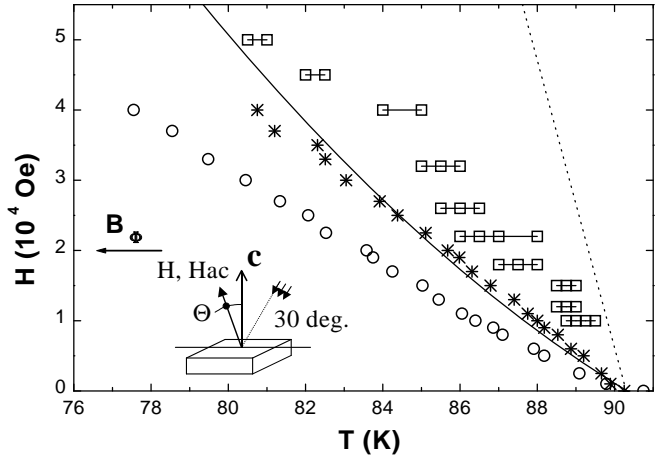}}
\caption{The irreversibility lines, circles and stars, obtained from
the onset of the screening component of the
{\it ac}--susceptibility with field oriented
at $\theta= -30^{\circ}$ and $\theta = +30^{\circ}$ with respect to the
{\it c}-axis.
The squares represent the conditions of temperature and magnetic field
under which the signature of the tracks could be observed in the reversible
torque signal.
Full and dotted lines are the melting line and $H_{c2}(T)$ obtained
in Refs.~\protect\onlinecite{7} and \protect\onlinecite{8}, rescaled by a
factor
$(\cos^2(\theta)+\gamma^{-2}\sin^2(\theta))^{-\frac{1}{2}}$
with $\theta =30^{\circ}$.
}
\label{irrline}
\end{figure}

In a separate experiment, the
irreversibility line was measured using SQUID {\it ac}-susceptometry.
It was located as the onset of the in--phase (reactive) component of
the {\em ac} susceptibility measured in an
oscillatory field of amplitude 0.1 Oe and frequency 13 Hz, oriented
parallel to the dc field.
These measurements were performed for two orientations of the static
field, applied parallel to the direction of the tracks (\em i.e. \rm
at 30$^{\circ}$ with respect to the $c$--axis), and applied in the
symmetric direction with respect to the $c$-axis. The irreversibility
fields for both orientations were found to be linear with
temperature; the line obtained with the field applied parallel to the
tracks clearly lies above the one for the symmetric orientation
(Fig.~\ref{irrline}). In contrast to what
is observed for Bi$_2$Sr$_2$CaCu$_2$O$_{8+\delta}$, and more recently,
in heavy-ion irradiated YBa$_2$Cu$_3$O$_{7-\delta}$ thick films,\cite{16}
there is no change in the behavior at $H  > B_{\phi}$ up to our
maximum measuring field of $H = 50$ kOe, and the lines do not merge above the
irradiation field.

\section{Results and Discussion}

\begin{figure}
\centerline{\epsfxsize=9.0cm\epsfbox{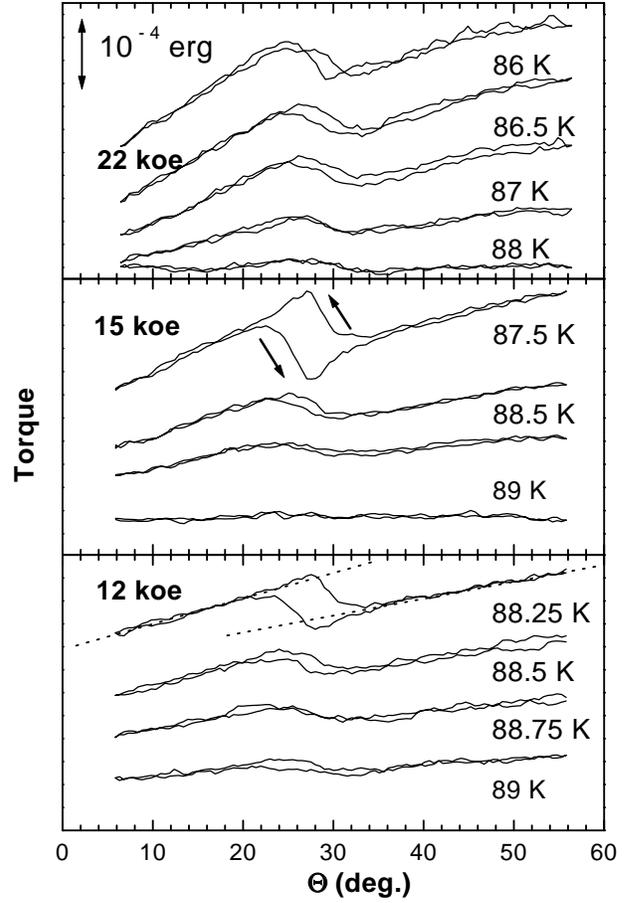}}
\caption{Torque signal in the vicinity of the irradiation direction
($\theta =30^{\circ}$). A smooth background has been subtracted
from the raw signal.
}
\label{data}
\end{figure}

Typical torque signals are displayed in Fig.~\ref{data}. Below the
irreversibility line determined by SQUID {\it ac}-susceptometry with
the field along the tracks,
the torque measurements reveal a hysteretic behavior when the field
is aligned with the irradiation direction. Above the line, the system
is in the so--called vortex liquid phase and the torque signal is reversible;
however, in a narrow region typically 1 to 2 K wide, a
kink is found, roughly symmetric with respect to the
orientation of the columnar defects (Fig.~\ref{data}).
This behavior is similar to what is observed
for conventional torque on a layered superconductor when the field
is rotated across the plane of the layers, and indicates that the
vortex lines deform in order to have their
direction coincide with that of the linear defects. In other words,
the free energy of the vortex liquid phase is lowered by flux--line
pinning onto the columnar tracks.

At low temperature, where thermal fluctuations are not important,
theory\cite{9}
predicts that when the external field is applied sufficiently close to the
layer/track direction (\em i.e. \rm the angle between applied field and the
tracks $\theta <\theta_l$ where $\theta_{l}$ is the lock-in angle),
the equilibrium configuration of a single flux-line is that in which the whole
length of the line is aligned with the defect. At larger angles
$\theta_l<\theta<\theta_t$, one expects a staircase configuration
in which line segments aligned with the defects alternate with segments
wandering between defects.\cite{9} For $\theta$ larger than the
accomodation angle $\theta_{t}$, the vortices do not readjust to the
columnar defects at all. In our experiment, it is unlikely
that we achieve the locked configuration, as this would require
the alignment of the external field with the track direction to within
some angle $\theta < \theta_{l} < \alpha$.
In the locked configuration, one should observe
a linear variation of the torque signal with angle, with a slope
$B^{2}/4\pi$ erg cm$^{-3}$ rad$^{-1}$ (neglecting the anisotropy in the
demagnetizing factors) i.e. $\approx  5 \times 10^{-2}$ erg deg$^{-1}$
in a 10 kOe field in our case. This is three orders of
magnitude larger than the highest of the
slopes in Fig.~\ref{data}.

\begin{figure}
\centerline{\epsfxsize=9.0cm\epsfbox{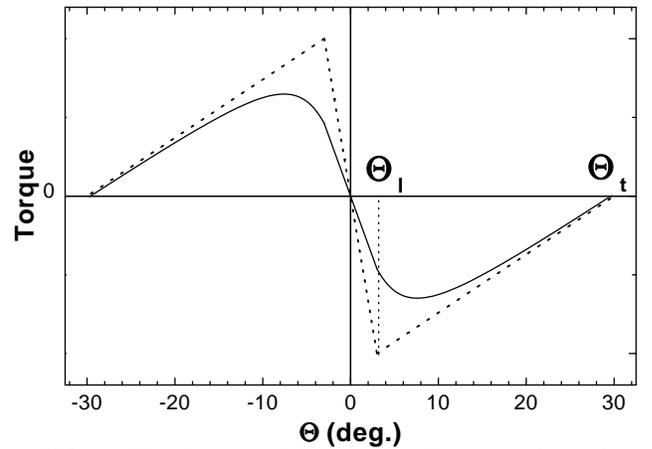}}
\caption{The theoretical torque signals arising from the interaction
of the vortices with the amorphous columnar defects
for two values of the misfit angle $\alpha$: solid
line $\alpha = 5^{\circ}$, dotted line $\alpha =0 $
The other parameters were $\theta_{l} = 3^{\circ}$ and $\theta_{t}
=30^{\circ}$.
}
\label{theory}
\end{figure}

The predicted contribution $\Gamma_{0}$ to the torque signal\cite{9}
is shown in Fig.~\ref{theory}.
As the field angle is increased from the irradiation direction, the torque
first increases linearly, reaching a maximum at the lock-in angle,
and then decreases linearly beyond this.
For angles larger than $\theta_t$, the torque contribution arising
from the interaction between vortices and ion tracks should be zero.
In practice, the lock-in angle is quite small, therefore
the torque signal should be quasi--discontinuous when the field-- and
track direction coincide. In the single vortex regime, the
magnitude of the torque signal close to the irradiation
direction may be obtained in terms of the lock-in angle,\cite{9}

\begin{eqnarray}
\Gamma_{0}(\theta) & = & \frac{B^{2}}{4\pi}  \theta_{l} \left(
\frac{| \theta |}{\theta_{t}} - 1 \right) {\mathrm sign}(\theta)
\label{eq:torque}
\\
\theta_{l} & = & \frac{4\pi \sqrt{2}\left(\varepsilon _{l}\varepsilon
_{r} \right)^{1/2}}{\Phi _{0}H},
\label{eq:lockinangle}
\end{eqnarray}

\noindent with $\varepsilon_r$ the pinning energy per unit length.
The accomodation angle $\theta _{t} =
\arctan(2\varepsilon _{r}/\varepsilon _{l})^{1/2}$, where
$\varepsilon_l=\gamma^{-2}\varepsilon_{0}\ln(1/K_{z}\xi)$
is the vortex line tension, the energy
scale  $\varepsilon_{0}=(\Phi_{0}/4\pi\lambda)^{2}$, and $K_{z}$ is the
typical wavevector of the vortex distortion induced by the columns.
In optimally doped YBa$_{2}$Cu$_{3}$O$_{7-\delta}$, the penetration depth
$\lambda = 1400 (1-T/T_{c})^{-1/2}$ \AA, the $ab$--plane coherence length
$\xi_{ab}(0) = 15$ \AA, and the anisotropy parameter $\gamma  \approx 7.6$.\cite{Paulius}
>From Eq.~(\ref{eq:torque}), one sees
that one can \em directly \rm obtain an estimate of the lock--in angle
from the torque jump observed when the field is aligned with the columns
and Eq.~(\ref{eq:torque}). A good estimate of the pinning energy
$\varepsilon_{r}$ is equally obtained from the torque jump:

\begin{equation}
	\varepsilon_{r} = \frac{(\Gamma_{0}a_{0})^{2}}{2\varepsilon_{l}}
	\label{eq:Upin}
	\end{equation}

\noindent ($a_{0} = (\Phi_{0}/B)^{1/2}$ is the mean
separation between vortices). This method to obtain the pinning
energy\cite{15} is more direct than estimates based on the angular dependence
of the resistivity $\rho$. Those rely on the identification of a shallow
maximum of $\rho$
at $\theta_{t}$, or, alternatively, with a
``depinning angle''\cite{10} determined by the rate at which vortices can
liberate themselves from a track; the relation of the latter with the
accomodation angle is not certain. Since the pinning
energy is predicted to be proportional to $\theta_{t}^2$, the method
based on transport measurements can result
in a large uncertainty in $\varepsilon_{r}$.

The present approach has
the advantage that there is only one assumption, which concerns the precise
form of
$\varepsilon_{l}$. Taking the curve at $H = 12$ kOe and $T =
88.5$ K ($T / T_{c} = 0.98$), \em i.e. \rm at the onset of magnetic
irreversibility, one has a typical value of the torque jump
$2\Gamma_0(0^-)\approx$ 400 erg cm$^{-3}$ (Fig.~\ref{data});
consequently, $\theta_l \approx 10^{-5}$ deg.
The parameter values $a_{0} = 400$ \AA, $\xi_{ab}(T)\approx 100$ \AA,
 $ K_{z} = 1 / a_{o}$, and
$ \lambda (T)\approx 7000 $ \AA, yield the pinning energy per unit
length $\varepsilon_{r}\approx 5 \times 10^{-9}$ erg cm$^{-1}$,
and $\theta_t \approx 70^{\circ}$.
The obtained value of the accomodation angle seems reasonable: the difference
between the extrapolation of the torque from large positive and negative angles to
$\theta = 0$, at which the field and the ion track are nearly
aligned, is a good indication that $\theta_{t}$ lies beyond the
angular range depicted in Fig.~\ref{data}. Clearly, $\theta_{t}$
greatly exceeds the angular width of the irreversible regime just below the
irreversibility line, which is about 8$^{\circ}$ at 88.5 K and $H = 12$ kOe.
The accomodation angle is comparable to the low--temperature limit of the
``depinning angle''
measured on an untwinned YBa$_{2}$Cu$_{3}$O$_{7-\delta}$ single crystal
irradiated with 1.0 GeV U ions to the same nominal dose.\cite{10}

Returning to the experimental data in Fig.~\ref{data},
one observes that, in spite of the fact that a clear jump in the
torque signal can be defined, the discontinuity at the irradiation
angle is rather smooth. The smoothness of the curve is possibly due to the
non-zero misalignment
angle $\alpha$. The effect of misalignment can be quantitatively accounted
for using simple
trigonometric considerations. Projecting the torque as given in
Ref.~\onlinecite{9} on the experimental torque axis {\bf u}
(Fig.~\ref{lever}), one obtains the magnitude of the measured torque
signal:

\begin{equation}
\Gamma=\frac{\sin(\theta)}{\sin(\tilde{\theta})}
\left|\Gamma_0(\tilde{\theta}) \right|
\end{equation}

\noindent where

\begin{equation}
\sin(\tilde{\theta}) =
\left[\sin^2(\theta)+\sin^2(\alpha)\cos^2(\theta)\right]^{1/2}.
\end{equation}

\noindent $\tilde{\theta}$ is the field rotation angle in the
laboratory frame, and $\theta$, as before, is the real angle between
the direction of the magnetic field and that of the ion tracks.
The curves plotted in Fig.~\ref{theory} shows that
the effect of the misalignment is  both to
widen the angular interval between the torque maxima
(now $\approx 2 \alpha$ from one another) and to decrease the torque
value at the maximum. Using $\theta_l\ll\theta_t$,
$\theta_t\approx30^{\circ}$, and $\alpha=5^{\circ}$
we find that the maximum torque is only about
0.6 $\Gamma_0(0^-)$ so that the pinning potential
estimated from the apparent torque jump is in this case only
about 40$\%$ of the actual value, which, at $H = 12$ kOe and
$T = 88.5$ K, would amount to $\varepsilon_{r} = 1.3 \times 10^{-8}$
erg cm$^{-1}$.

The absolute value of the pinning energy per unit length is in
reasonable agreement with the estimate for core pinning of individual
vortices,\cite{Nelson93}

\begin{equation}
	\varepsilon_{r} = \varepsilon_{0} \left( \frac{c_{0}}{2\xi_{ab}}
\right)^{2} \propto (1-T/T_{c})^{2}
	\label{eq:core}
	\end{equation}

\begin{figure}
\centerline{\epsfxsize=9.0cm\epsfbox{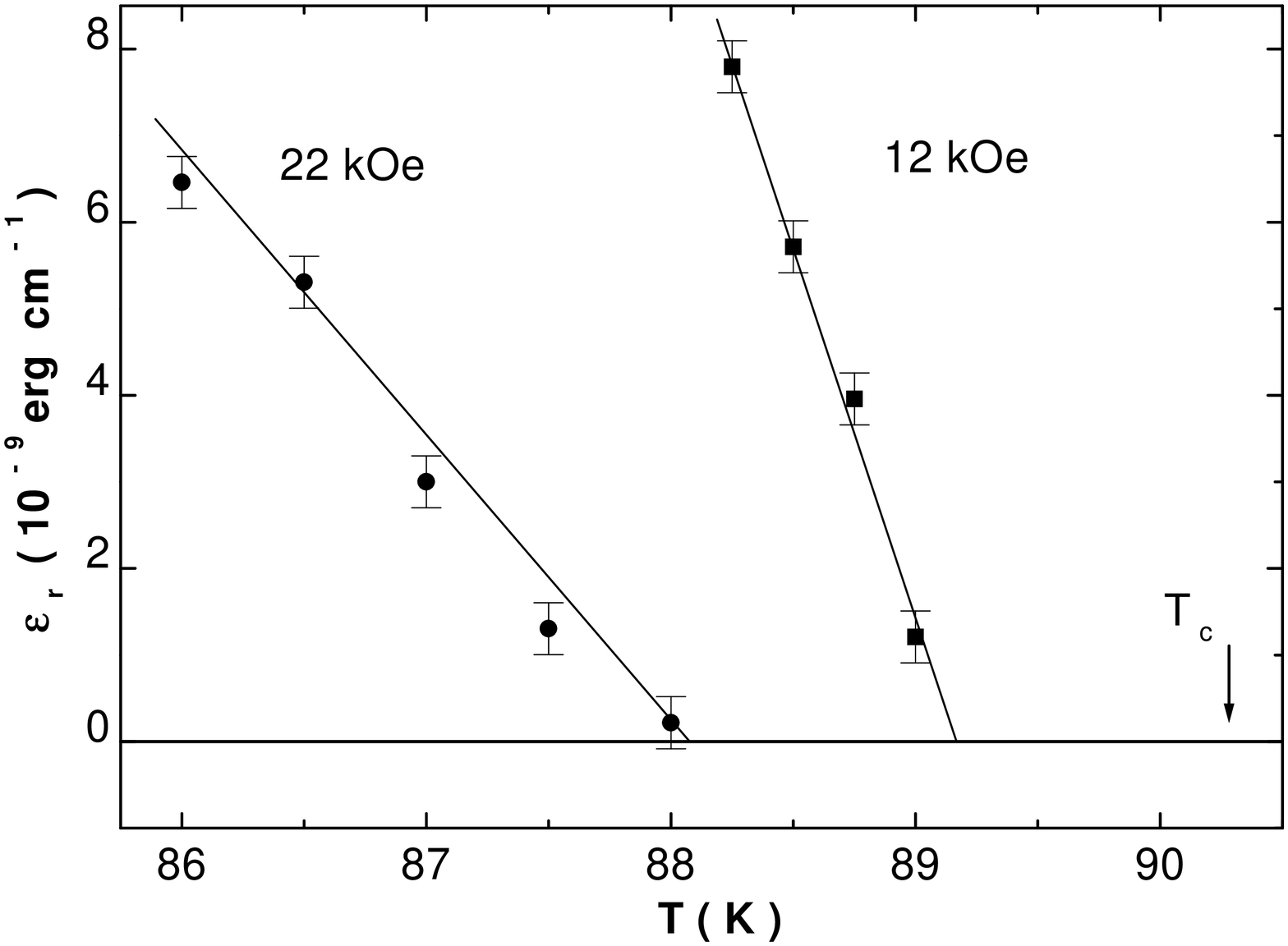}}
\caption{
The pinning energy, as given by the torque jump using
Eq.~\protect\ref{eq:Upin} and the parameters given in the text.
Drawn lines are guides to the eye.
}
\label{fig:Upin}
\end{figure}

\noindent (with $\varepsilon_{0}(0) = 5 \times 10^{-8}$ erg cm$^{-1}$)
In the temperature regime of interest, this mechanism
is more relevant than electromagnetic pinning\cite{11} because
$\xi_{ab}(T)$ greatly exceeds the track radius. Using the same parameter
values as above, the model yields the theoretical value $\varepsilon_{r}
\approx 1 \times 10^{-9}$
erg cm$^{-1}$ (at $T/T_{c} = 0.98$).  Recent measurements on
heavy--ion irradiated Bi$_{2}$Sr$_{2}$CaCu$_{2}$O$_{8+\delta}$\cite{Drost99}
showed that in that material, the dependence of pinning energy on
track diameter and temperature is in agreement with the core pinning
model, although the magnitude of the pinning energy exceeded the
theoretical expectation (\ref{eq:core}) by a factor 5. In the
present case, the strong temperature and field dependence of the
experimentally obtained
pinning energy, displayed in Fig.~\ref{fig:Upin},
show that a simple ``zero temperature'' single vortex pinning approach is
inadequate.
The reasons for this are that (i) the fields under consideration are not
small with respect to
$B_{\phi}$, so that only a fraction of vortices can be expected to be actually
trapped on a columnar track, and (ii) the proximity to $T_{c}$
possibly necessitates the inclusion of the effect of strong thermal
fluctuations.\cite{Nelson93,Samoilov96}

A theoretical description of the effect of a field rotation, or even
of the total pinning energy, in the case where the vortex density is
comparable to the density of a system of strong linear pins has,
to our knowledge, not been developped at present. Although the decrease of
the pinning energy per unit volume as field is increased, and the eventual
disappearance of the torque jump at $H \gg B_{\phi}$, is the
straightforward consequence of the averaging of the pinning energy
gain obtained from the restricted number of vortices trapped on an ion
track and the ever increasing number of those that are not, there are
few predictions about the resulting field dependence of the
equilibrium magnetization. Extensive numerical calculations of the
vortex energy distribution in the presence of columnar pins were
carried out by Wengel and T\"{a}uber;\cite{Wengel98} however, they did
not make any specific predictions as to the precise temperature or
field dependence of the magnetization.

The effect of thermal fluctuations must also be considered. In resistivity
measurements, such fluctuations are usually accounted for by
stating that, at the angle at which depinning occurs, \em i.e. \rm at
which the probability to find a pinned vortex segment becomes
exponentially small, the thermal energy
and the pinning energy of a single trapped vortex segment
are equal.\cite{10,13} As a consequence, the ``depinning angle''
measured by the angular dependence of the resistivity is smaller
than the accomodation angle and is given by:\cite{10}

\begin{equation}
\varepsilon_{r}\approx \varepsilon_{0}\left(\frac{2k_{B}T\tan \theta
_{depin}}{\varepsilon _{0} a_{0}} \right)^{2/3}.
\end{equation}

\noindent With the parameters values as used above, we obtain $
\theta_{depin}\approx $
15$^{\circ}$, which is comparable to the angular width of the irreversible
regime in Fig.~\ref{data}. The same type of argument leads one to
conclude that at the same temperature, thermal fluctuations are much
less efficient when the field is aligned with the track direction,
because the length and the trapping energy of the
pinned line segments are large. Nevertheless, the fact that the
magnetization is \em reversible \rm and that the resistivity measured under
similar condictions is \em linear \rm\cite{13} implies that although it may
be small, the thermal depinning rate is non--zero. Since the measured
pinning energy is proportional to the average vortex length trapped on
a columnar defect at any one moment, the rapid decrease of
$\varepsilon_{r}$ with temperature, which is in agreement with estimates
obtained from
resistivity data,\cite{10} does not seem to be an artefact of the method
used to analyze torque or resistivity data, but reflects the
increasing efficiency of thermal fluctuations in liberating vortices
from the tracks.  Strong vortex
wandering above a ``depinning temperature'' $T^{*} =
(c_{0}/\pi)\sqrt{\varepsilon_{l}\varepsilon_{r}}$,
such as proposed in Refs.~\onlinecite{9,Nelson93}, and \onlinecite{Samoilov96}
would lead to a torque jump that follows an exponential temperature
dependence.  Such a dependence was observed in Ref.~\onlinecite{17},
where the rapid decrease of the accomodation angle measured
in heavy-ion irradiated Bi$_2$Sr$_2$CaCu$_2$O$_{8+\delta}$
was attributed to the effect of thermal fluctuations. Although the reduced
range of temperatures over which $\varepsilon_{r}$ could be determined
in the present experiments makes a direct comparison very
difficult, thermal wandering of flux lines could be responsible for
the disappearance of the pinning energy and the torque jump at
temperatures \em below \rm $T_{c2}(H)$ (see Fig.~\ref{irrline}).

\section{Conclusion}
We have, from thermodynamic torque measurements,
obtained the first evidence for an angle--dependent contribution
of amorphous columnar defects to the equilibrium magnetization in
YBa$_{2}$Cu$_{3}$O$_{7-\delta}$. The analysis of the torque signal
allowed us to directly determine the lock-in angle and the pinning
energy of the linear defects. The magnitude of the pinning energy is in
qualitative agreement with the core pinning mechanism by columnar
defects; however, the observed strong field dependence means that the
interactions between flux lines are not negligible in the range of
magnetic fields investigated here. The strong temperature dependence
of the torque jump, and the disappearance of the pinning energy below
$H_{c2}(T)$ are the consequence of thermal fluctuations in the vortex
liquid state, which are increasingly efficient in liberating vortex
segments from the tracks as temperature increases.

\section{Acknowledgements} The work of STJ is funded by the EC, TMR grant Nr.
ERBFMBICT961728. We thank F. Holtzberg (emeritus, I.B.M. Thomas J.
Watson Research Center, Yorktown Heights) for providing the
YBa$_{2}$Cu$_{3}$O$_{7-\delta}$ single crystal.

\newpage

\end{multicols}

\begin{references}
\bibitem{Paulius} For the sake of comparison, we use the same
parameter values as Ref.~\protect\onlinecite{10}.
\end{references}
\end{document}